\documentclass{Interspeech}

\interspeechcameraready
\usepackage{comment}
\usepackage{multirow}
\usepackage{booktabs}
\usepackage{tablefootnote}

\title{Adapting Whisper for Streaming Speech Recognition via Two-Pass Decoding}

\author[affiliation={1,2}]{Haoran}{Zhou}
\author[affiliation={2}]{Xingchen}{Song}
\author[affiliation={1}]{Brendan}{Fahy}
\author[affiliation={2}]{Qiaochu}{Song}
\author[affiliation={2}]{Binbin}{Zhang}
\author[affiliation={2}]{Zhendong}{Peng}
\author[affiliation={1}]{Anshul}{Wadhawan}
\author[affiliation={1}]{Denglin}{Jiang}
\author[affiliation={1}]{Apurv}{Verma}
\author[affiliation={*1}]{Vinay}{Ramesh}
\author[affiliation={1}]{Srivas}{Prasad}%
\author[affiliation={1}]{Michele M.}{Franceschini}

\affiliation{}{Bloomberg}{USA}
\affiliation{}{WeNet Open Source Community}{China}
\email{hzhou245@bloomberg.net, vramesh7@bloomberg.net}

\newcommand\blfootnote[1]{%
  \begingroup
  \renewcommand\thefootnote{}\footnote{#1}%
  \addtocounter{footnote}{-1}%
  \endgroup
}
\begin{document}

\maketitle

\begin{abstract}
OpenAI Whisper is a family of robust Automatic Speech Recognition (ASR) models trained on 680,000 hours of audio.
However, its encoder-decoder architecture, trained with a sequence-to-sequence objective, lacks native support for streaming ASR.
In this paper, we fine-tune Whisper for streaming ASR using the WeNet toolkit by adopting a Unified Two-pass (U2) structure.
We introduce an additional Connectionist Temporal Classification (CTC) decoder trained with causal attention masks to generate streaming partial transcripts, while the original Whisper decoder reranks these partial outputs.
Our experiments on LibriSpeech and an earnings call dataset demonstrate that, with adequate fine-tuning data, Whisper can be adapted into a capable streaming ASR model.
We also introduce a hybrid tokenizer approach, which uses a smaller token space for the CTC decoder while retaining Whisper’s original token space for the attention decoder, resulting in improved data efficiency and generalization.
\end{abstract}

\section{Introduction}
Large-scale training has provided substantial improvements in both accuracy and robustness for speech recognition models.
Whisper \cite{radford2023robust}, released by OpenAI, exemplifies this trend.
It was trained on 680,000 hours of audio and achieves high performance across diverse public benchmarks.
However, due to its non-causal design, it inherently lacks support for streaming speech recognition.
\blfootnote{* Corresponding author}
Multiple efforts have aimed to adapt Whisper for streaming ASR.
The UFAL streaming Whisper implementation \cite{machavcek2023turning} buffers incoming audio until it reaches a designated length before running the Whisper model,
then uses a streaming policy that compares consecutive non-streaming predictions to finalize transcripts.
Another approach, Simul-Whisper \cite{wang24ea_interspeech}, further employs input truncation to avoid splitting word boundaries and improve pseudo-streaming inference.
Neither of these methods requires Whisper fine-tuning;
however, neither fully converts Whisper into a true streaming model because they still rely on Whisper’s original non-streaming interface.
Running inference on incomplete audio segments with a purely sequence-to-sequence training objective leads to a mismatch between training and inference, hampering accuracy in low-latency scenarios.
Furthermore, inference efficiency is suboptimal, as these approaches repeatedly process the same audio content.
Additionally, Whisper models are trained with input padded to 30 seconds.
Because of this, the inference stage has to pad inputs to 30 seconds regardless of their actual length, adding extra computational overhead.

By contrast, most streaming ASR models enforce high-quality partial transcripts on partial inputs through alignment-free training.
For example, CTC \cite{graves2006connectionist} and Recurrent Neural Network Transducer (RNN-T) \cite{graves2012sequencetransductionrecurrentneural} architectures are natively well-suited for streaming because they achieve causality through implicit alignment in training.
A U2 model \cite{zhang2020unified}, implemented in WeNet \cite{yao2021wenet} \cite{binbin2022wenet2}, provides a natural solution to adapt Whisper for streaming speech recognition, as it combines an encoder-decoder architecture with streaming capabilities.

We propose U2 Whisper, which adapts Whisper for streaming ASR under the U2 model architecture.
Specifically, we introduce a CTC decoder on top of the Whisper encoder and fine-tune the model with a hybrid CTC-attention loss.
The CTC branch is trained for streaming recognition, while the original Whisper decoder branch is retained for the original sequence-to-sequence setup.
During inference, the CTC decoder provides streaming partial hypotheses.
Upon detecting an endpoint, the Whisper decoder is used for rescoring to determine the best final result.
These adaptations not only turn Whisper into a causal streaming model, but also improve efficiency during inference, allowing the model to run on CPUs in real time.

Nevertheless, our ablation studies reveal that the U2 path alone may not generalize well to challenging test sets if only limited data is used for fine-tuning.
Hence, we reduce the token space for the CTC decoder, leveraging more granular subword token modeling to better handle rare words, especially in low-resource conditions \cite{sennrich2016neural}.
We do so by selecting the first 8,000 tokens from Whisper’s original tokenizer for the CTC decoder while leaving the original full-token tokenizer intact for the Whisper decoder.
Experimental results demonstrate that this hybrid-tokenizer method improves generalization, particularly when training data is scarce.

\section{Methods}

\subsection{Adapting Whisper to a U2 model}
\begin{figure}[htbp]
  \centerline{\includegraphics[width=8.5cm]{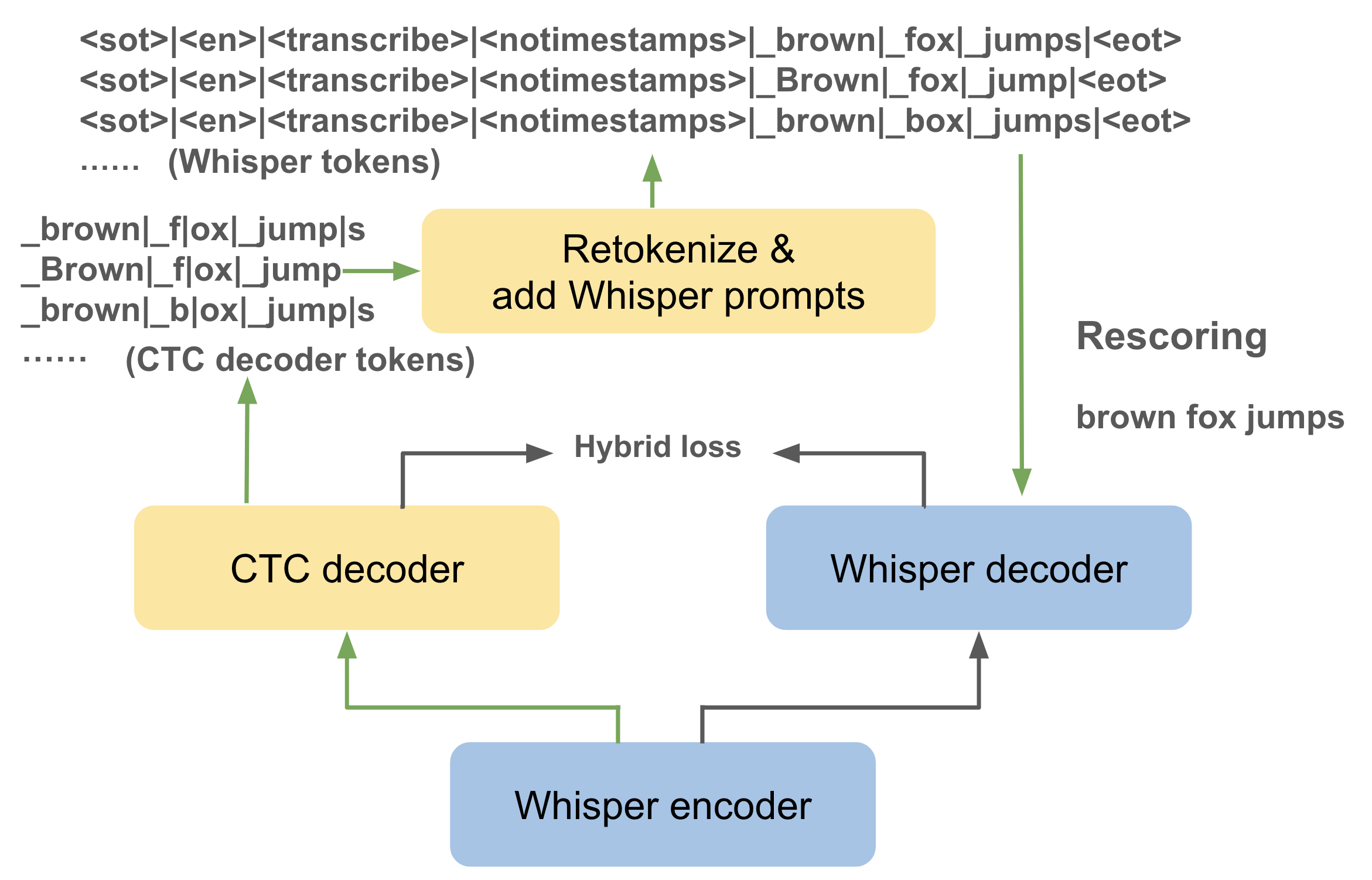}}
  \caption{Streaming Whisper with a hybrid tokenizer using a Unified Two-pass decoding framework.}
  \label{figure:u2-whisper}
\end{figure}

U2 ASR models \cite{zhang2020unified} aim to provide a unified architecture for both non-streaming and streaming ASR.
The U2 model comprises an encoder, a CTC decoder, and an attention decoder.
Figure~\ref{figure:u2-whisper} illustrates how we adapt Whisper using the U2 structure.

During training, both the CTC and attention decoders learn to generate the reference transcript using a hybrid CTC-attention loss \cite{watanabe2017hybridloss}, as shown in Equation~\ref{equation:hybrid_loss}.
\begin{align}
  \mathcal{L} = \alpha \cdot \mathcal{L}_{\text{CTC}} + (1 - \alpha) \cdot \mathcal{L}_{\text{Attention}}
  \label{equation:hybrid_loss}
\end{align}

U2 also introduces dynamic attention masks during training to ensure the encoder’s hidden representations depend only on past or a small portion of future context.
Training with these attention masks enables the encoder to run in streaming mode at inference time, ensuring consistent behavior between training and streaming inference.
In our experiments, we sample chunk sizes randomly between 0.1 and 1.0 seconds during training, helping the model generalize across a range of chunk sizes.

\subsection{Streaming inference}
\label{sec:streaming-inference}
Inference flow is illustrated by the green arrows in Figure~\ref{figure:u2-whisper}.
The encoder processes audio input in chunks, and the CTC decoder performs prefix beam search \cite{graves2006connectionist} to produce the top-k streaming partial transcripts.
When an endpoint is detected, either after 0.5 seconds of silence or when the max delay constraint is reached, the attention decoder's rescoring pass finalizes the partial transcript.
The final transcript is selected by rescoring these top-k CTC hypotheses and choosing the highest-scoring one.

All evaluations are conducted using the WeNet C++ inference runtime, where our open source implementation of Whisper inference support is available.
This setup enables us to test production-time performance in an end-to-end manner,
particularly for long-form streaming transcription. 
Such comparisons offer better insights into the model's performance compared to the original Whisper model, which benefits significantly from the long historical transcript used as a prompt during long-form transcription.

The WeNet toolkit implements an efficient key-value (KV) cache for incremental streaming encoder inference,
allowing reuse of KV values from previous chunks without recomputation.
During attention rescoring, the system only needs a single batched pass through the decoder with a diagonal causal attention mask, without the need for autoregressive decoding.
These optimizations improve inference efficiency and allow real-time CPU-based processing for fine-tuned Whisper Medium models,
despite their substantial size of 769 million parameters.

\subsection{Hybrid tokenizer}
Whisper uses a large token space derived from the GPT-2 tokenizer, based on Byte Pair Encoding (BPE) \cite{gage1994new}, totaling over 50,000 tokens \cite{radford2019language}.
When Whisper is fine-tuned on a small, domain-specific dataset, this extensive token space may be insufficiently covered for training the CTC decoder effectively.
As a result, the CTC branch, trained from scratch, may struggle with out-of-domain or rare tokens, leading to poor generalization.

To address this, we restrict the token space for the CTC decoder to the first 8,000 tokens of the Whisper tokenizer.
This ensures coverage of essential components such as numbers, uppercase and lowercase letters, and common subwords, enabling efficient tokenization without sacrificing performance.
During training, we use these 8,000 tokens to form the CTC prediction targets via SentencePiece \cite{kudo2018sentencepiecesimplelanguageindependent}, while the attention decoder maintains the full token set.

For inference, we implement a TorchScript-compatible retokenizer with TorchText in PyTorch \cite{AnselPyTorch2Faster2024}.
The retokenizer decodes the CTC hypotheses into strings.
It then retokenizes them with the Whisper tokenizer and adds Whisper-specific prompt tokens before passing the sequences to the attention decoder for rescoring, as shown in Figure~\ref{figure:u2-whisper}.

\section{Datasets}
Although we include experiments on LibriSpeech \cite{panayotov2015librispeech} for completeness and comparison with other approaches,
our primary focus is on an internally curated dataset of earnings calls.
We also considered Earnings-22 \cite{del2022earnings}, but excluded it due to its limited size, which is insufficient for training required by the experiments.
The internal earnings dataset was selected for its high-quality, written-form transcripts, which incorporate proper punctuation, capitalization, and formatting (e.g., for emails and numbers), aligning closely with Whisper’s written-form outputs.
By centering on this dataset, we aim to fine-tune Whisper into a streaming ASR model that can provide transcription, punctuation, capitalization, and inverse text normalization in a unified manner.

Additionally, the dataset serves as a challenging test of generalization, given its dense financial terminology that is often specific to individual earnings calls.
This allows us to evaluate the model’s ability to handle domain-specific vocabulary and unseen words.
Compared to LibriSpeech, the earnings calls also represent a more realistic production environment, featuring relatively long audio samples that test long-form transcription, natural pauses, and a moderate degree of noise.
Another advantage is that we have ample high-quality data, allowing us to investigate how the model's performance saturates as the amount of training data increases.

We use randomly sampled earnings call data from before 2023 as our training set, yielding 5,800 hours of audio with accompanying text transcriptions, segmented into 5 to 20-second clips using a forced aligner.
To prevent data leakage, our test set is sampled from 83 calls that took place after 2023, resulting in 83 samples with a total duration of 10 hours.
This setup provides representative coverage of the target distribution, while ensuring that the test samples are long enough to evaluate end-to-end streaming performance.

\section{Results}
To assess how our proposed approach scales with data,
we fine-tune the Whisper Medium model using subsets of the earnings call training data.
The subsets consist of 725, 1,450, 2,900, and 5,800 hours of audio, respectively, and we evaluate both the single-tokenizer and hybrid-tokenizer approaches.
We first train with causal attention masks, focusing solely on the attention loss for one epoch to adapt Whisper to inputs shorter than 30 seconds and ensure the encoder is streaming-compatible.
Next, we add the CTC classification head and train for two epochs using only the CTC loss, keeping all other parameters frozen.
Since loss saturation with respect to training steps is typically observed after just the first epoch, training with one or two epochs is sufficient for those two steps.
Finally, we unfreeze all parameters and apply a hybrid loss until validation Word Error Rate (WER) stops improving for three consecutive epochs.
These procedures keep the parameters close to the pretrained model for better generalization.

In all U2 Whisper experiments, we set the prefix beam search size to 10 for the CTC decoder, select the top 6 candidate hypotheses for rescoring, and employ 8-bit quantization to reduce streaming latency.
The quantized model results in less than 0.3\% absolute WER degradation, so it does not affect our overall findings.
All evaluations use a 1-second chunk size and a 12-second maximum delay unless otherwise stated. No language model is used.

\subsection{Data scaling behavior of this approach}
Table \ref{table:training-size} presents WER results on the earnings test set for both single-tokenizer and hybrid-tokenizer setups across different training set sizes.
The hybrid tokenizer consistently performs better, especially when training data is limited, but its advantage decreases as the dataset size grows.
Training the same Whisper Medium model structure from scratch using the full training set, without the pretrained Whisper weights, yields a WER of 20.59\%, compared to 17.30\% obtained with the pretrained model, confirming the value of the pretrained weights.

\begin{table}[th]
  \caption{WER of fine-tuned streaming Whisper Medium with varying sizes of training data}
  \label{table:training-size}
  \centering
  \begin{tabular}{ccc}
    \toprule
    \multirow{2}{*}{\textbf{Training Set Size}} 
      & \multicolumn{2}{c}{\textbf{WER}} \\
    \cmidrule(lr){2-3}
      & \textbf{Single Tokenizer} 
      & \textbf{Hybrid Tokenizer} \\
    \midrule
    725\,h  & 23.51\% & 21.09\% \\
    1450\,h & 20.78\% & 18.97\% \\
    2900\,h & 19.67\% & 18.26\% \\
    5800\,h & 17.51\% & 17.30\% \\
    \bottomrule
  \end{tabular}
\end{table}

\subsection{Runtime configurations and performance}
\label{ssec:runtime-config}
We evaluate the best-performing checkpoint, which is trained with 5,800 hours, under different chunk sizes, with results on the earnings test set shown in Table \ref{table:chunk-size}.

\begin{table}[th]
  \caption{WER of fine-tuned streaming Whisper Medium using various chunk sizes, with and without rescoring}
  \label{table:chunk-size}
  \centering
  \begin{tabular}{ccc}
    \toprule
    \multirow{2}{*}{\textbf{Chunk Size}} 
      & \multicolumn{2}{c}{\textbf{WER}} \\
    \cmidrule(lr){2-3}
      & \textbf{w/o Rescoring} 
      & \textbf{w/ Rescoring} \\
    \midrule
    100\,ms  & 26.93\% & 25.54\% \\
    240\,ms  & 21.53\% & 21.20\% \\
    500\,ms & 18.67\% & 18.35\% \\
    1000\,ms & 17.60\% & 17.30\% \\
    1500\,ms & 16.85\% & 16.65\% \\
    \bottomrule
  \end{tabular}
\end{table}

Accuracy declines as the chunk size decreases.
We observe more formatting errors with smaller chunks (e.g., “\$1.3 million” might become “1.3 million dollars”),
because relevant future context needed for correct formatting is missing at the time of partial decoding.
If the correct hypothesis is pruned early during CTC prefix beam search, subsequent steps cannot fix the formatting.

Rescoring does improve accuracy by leveraging the attention decoder’s knowledge to select the best hypothesis,
but Table \ref{table:chunk-size} shows that the effect is relatively modest. Typically, the top CTC-generated hypotheses differ only in small details, such as punctuation or capitalization, limiting the potential benefit of rescoring.

The maximum delay parameter also significantly affects performance.
Table~\ref{table:finalize-latency} reports the WER, Real-Time Factor (RTF), and average finalize latency—measured as the average time taken by the finalize operation (i.e., attention rescoring)—under different maximum delay settings.
All U2 experiments are done with four virtual CPU cores (Intel Xeon 6240).

\begin{table}[htbp]
  \centering
  \caption{Performance of the model with different finalize latency constraints}
  \label{table:finalize-latency}
  \begin{tabular}{cccc}
    \toprule
    \multirow{2}{*}[0pt]{\textbf{Max Delay}}
    & \multirow{2}{*}[0pt]{\textbf{WER}}
    & \multirow{2}{*}[0pt]{\textbf{RTF}}
    & \multirow{2}{*}{\shortstack{\textbf{Average Finalize}\\ \textbf{Latency}}} \\
    & & & \\
    \midrule
    8 s  & 19.26\% & 0.23 & 679 ms \\
    12 s & 17.30\% & 0.30 & 1126 ms \\
    16 s & 17.23\% & 0.32 & 1490 ms \\
    20 s & 16.96\% & 0.34 & 1935 ms \\
    \bottomrule
  \end{tabular}
\end{table}

While longer delays help improve WER, they also increase computational cost because computational complexity grows quadratically with input length.
Selecting the maximum delay is crucial for balancing accuracy and runtime efficiency.

When considering end-to-end latency, three main components come into play:
chunk-based buffering latency (excluding computation),
partial transcript computation (around 267 ms in our setup),
and finalize computation latency (see Table \ref{table:finalize-latency}).
In our experiments, the U2 Whisper model can run with acceptable streaming latencies at a 12-second maximum delay.
Nonetheless, finalize computation latency remains relatively high for real-time applications, and can be reduced using Whisper Turbo checkpoints because of the smaller decoder.

\subsection{Comparing with UFAL Whisper and other variants on earnings and LibriSpeech}
\label{ssec:compare}
We compare the U2 streaming Whisper approach against UFAL Whisper and other non-streaming/offline models on both the earnings and LibriSpeech datasets.
To ensure a fair comparison, we first fine-tune Whisper Medium following the same setup described in the Whisper paper \cite{radford2023robust}; then load the resulting weights into UFAL inference code.
In UFAL inference, we use a beam size of 5 by default, along with all other parameters set to default values, running on the Faster Whisper back-end with an A100 GPU and sufficient CPUs, while all U2 models run on 4 Xeon 6240 CPU cores.
Additional non-streaming Whisper baselines are also included, with “FT” indicating models fine-tuned on the corresponding training set.

\begin{figure}[t]
  \centering
  \includegraphics[width=7.5cm]{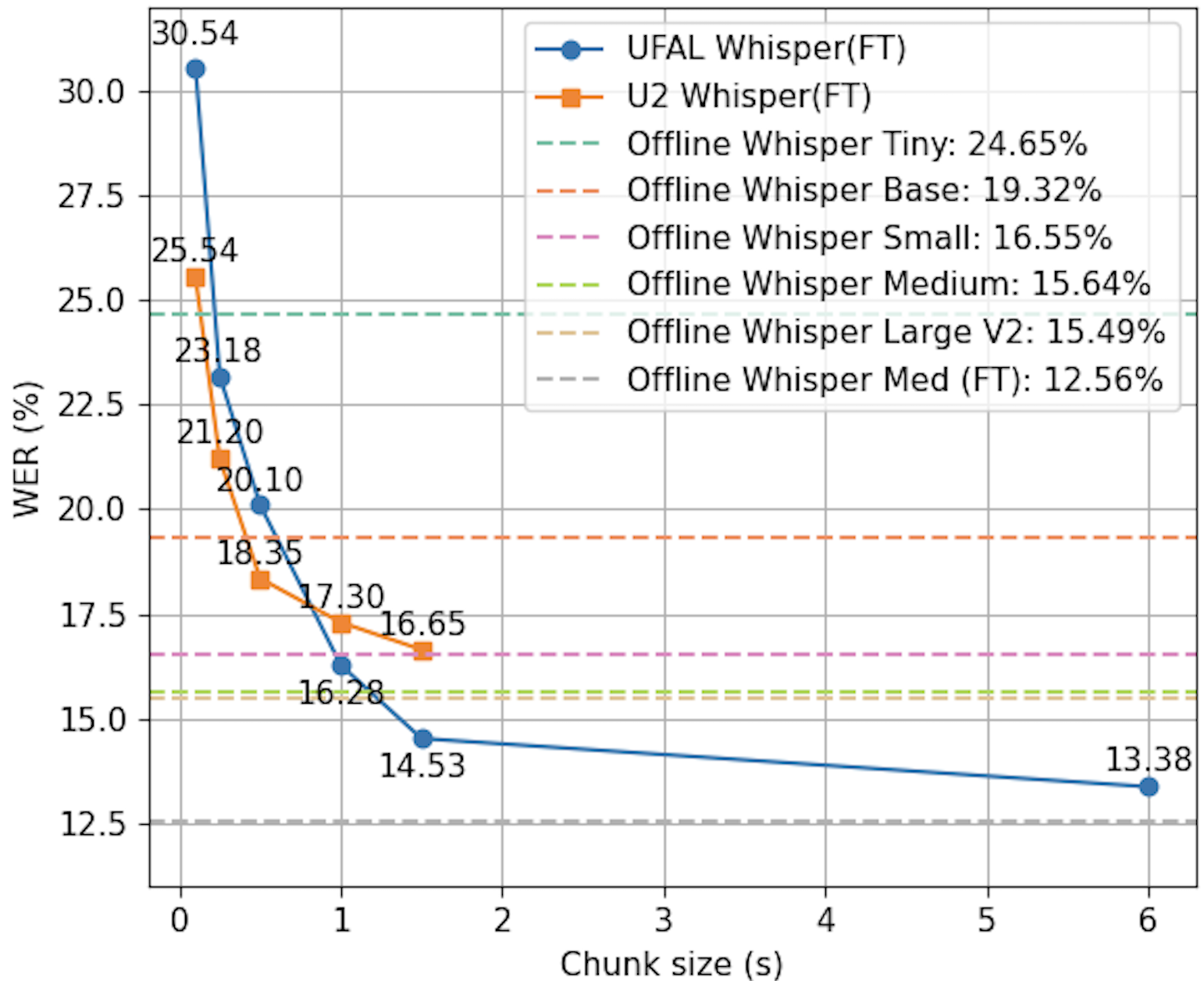}
  \caption{WER with different chunk sizes on the earnings test set.}
  \label{figure:earnings}
\end{figure}

Figure~\ref{figure:earnings} shows results on the earnings test set using chunk sizes of 0.1, 0.24, 0.5, 1, 1.5, and 6 seconds.
We do not report an offline U2 Whisper WER here because the model cannot perform offline inference on the lengthy test samples without segmentation.

\begin{figure}[t]
  \centering
  \includegraphics[width=7.5cm]{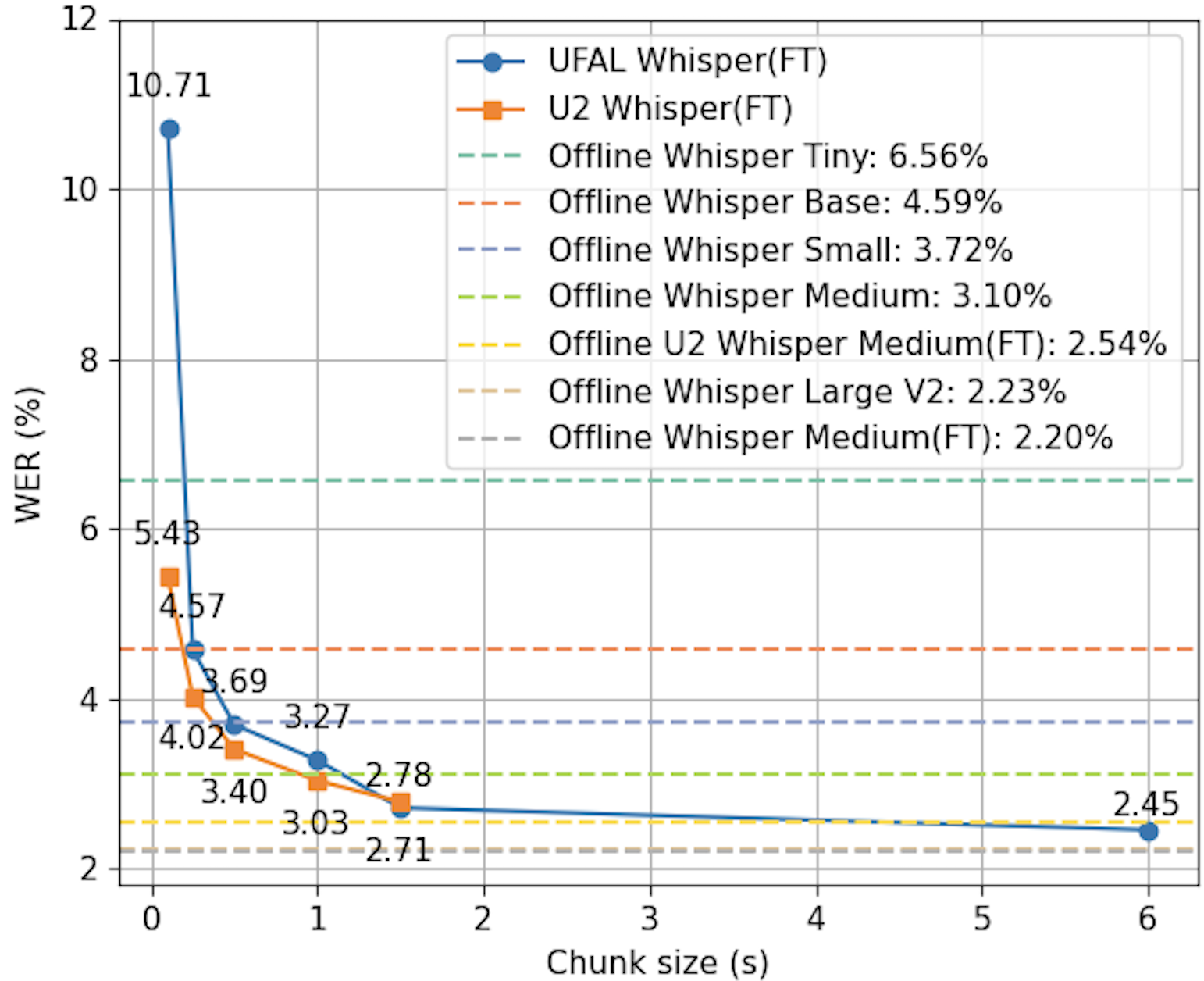}
  \caption{WER with different chunk sizes on LibriSpeech test-clean.}
  \label{figure:librispeech-clean}
\end{figure}

\begin{figure}[t]
  \centering
  \includegraphics[width=7.5cm]{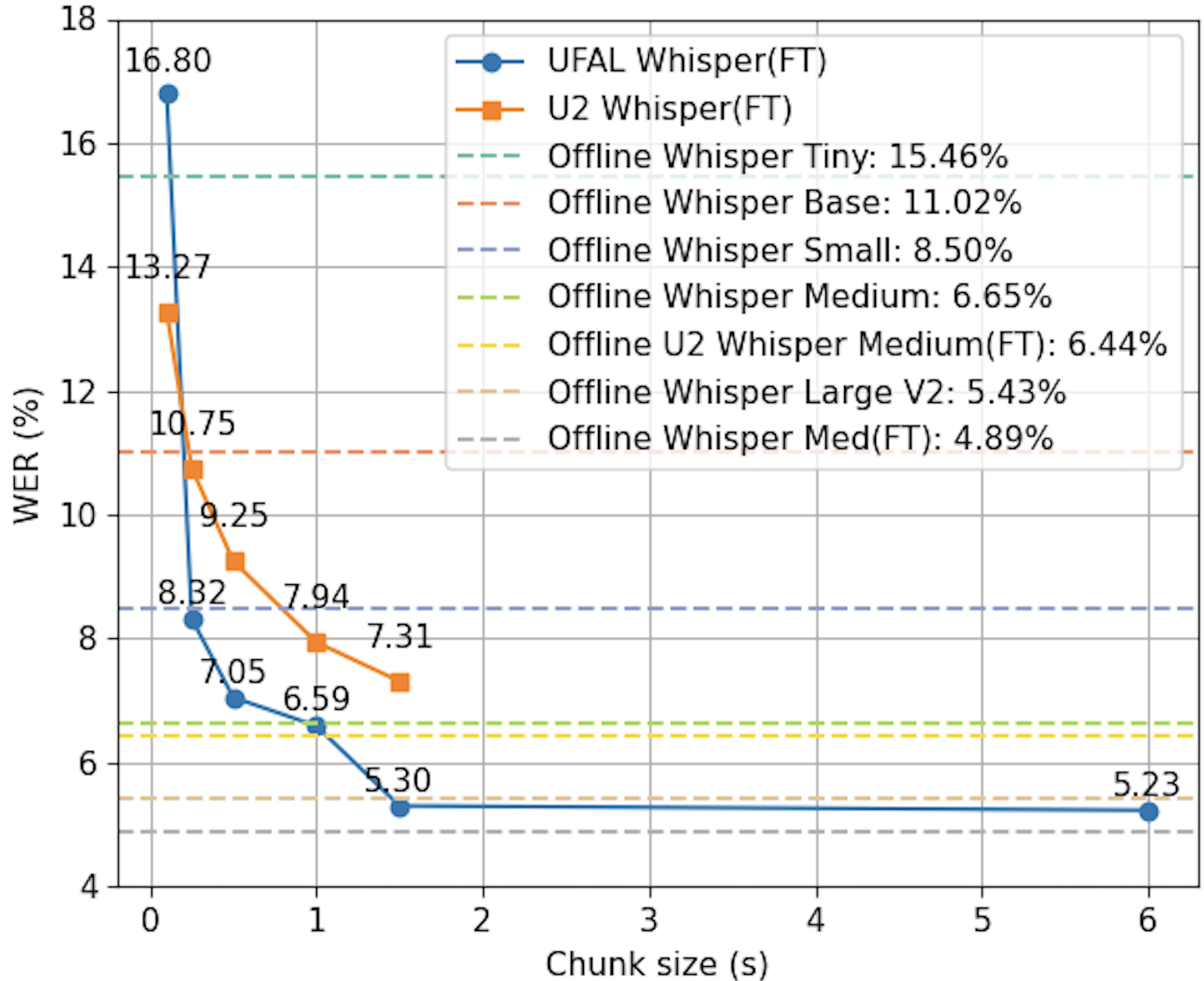}
  \caption{WER with different chunk sizes on LibriSpeech test-other.}
  \label{figure:librispeech-other}
\end{figure}

We conduct a similar evaluation on LibriSpeech,
training only on its standard training partition.
Figures~\ref{figure:librispeech-clean} and \ref{figure:librispeech-other} show results on test-clean and test-other, respectively.
Training a model from scratch with the same structure but without using the Whisper Medium weights yields a 5.18\% WER on LibriSpeech test-clean and 13.35\% on test-other.

UFAL’s original implementation does not emit unconfirmed partial transcripts for each chunk;
it emits final transcripts only when two consecutive predictions match.
In practice, it is possible to modify UFAL to produce partial transcripts.
From this perspective, testing both models with the same chunk size roughly corresponds to the same computation-unaware partial transcript emission latency.
In practice, forcing immediate partial output in UFAL can lead to hallucinations in partial outputs,
partly explaining the WER degradation with smaller chunk sizes.
However, assessing partial transcript quality is challenging because word boundaries can span multiple chunks, and future updates may overwrite earlier outputs,
so we are only comparing WER on finalized/confirmed transcripts.

For small chunk sizes, U2 Whisper outperforms UFAL Whisper on the earnings set and on LibriSpeech test-clean.
On the more challenging LibriSpeech test-other, similar to the earnings scenario, more training data is needed for U2 Whisper to surpass UFAL.
When the chunk size grows larger, UFAL Whisper has an advantage, likely because it more closely resembles non-streaming mode.

Regarding computational efficiency, U2 runs efficiently on CPUs, while UFAL requires more computation.
UFAL does not achieve real-time speed with Whisper Medium on CPUs, even with 8-bit quantization.
On GPUs, UFAL still struggles to match real time, i.e. RTF $>$ 1, if the chunk size is 0.5 seconds or lower for LibriSpeech, or 1.0 seconds for the earnings data.

In terms of computation-unaware finalize latency, UFAL finalizes words once two consecutive predictions match, typically corresponding to twice the chunk size \cite{machavcek2023turning}.
If the chunk size is 1.0 seconds, UFAL tends to have an average final emission latency of about 2.0 seconds.
By contrast, U2 Whisper only finalizes partial transcripts upon endpoint detection, usually during pauses,
which may result in a higher average final emission latency.  
However, the maximum delay parameter in U2 Whisper provides a guaranteed upper limit on how long the model will wait before forced finalization,
while the upper bound of UFAL’s final transcript emission latency is not strictly bounded.

In summary, when enough in-domain data is available, U2 Whisper is better suited for real-time applications where low-latency partial transcripts are essential and computational efficiency is a priority.
On the other hand, UFAL Whisper is more suitable for scenarios where immediate partial transcripts are less critical, and where abundant GPU resources are available.

\section{Conclusion}
\label{sec:conclusion}
We have presented a method for adapting Whisper into a streaming ASR model using the U2 architecture, achieving performance comparable to the original Whisper setup.
We introduced a hybrid tokenizer to improve generalization, especially when fine-tuning is performed with limited data.
Our experiments highlight the data-scaling behavior of this approach and show that it can run in real-time on CPUs under appropriate runtime configurations.

We observed that the hybrid tokenizer yields substantial gains in low-resource conditions but offers diminishing returns with larger datasets.
We also examined various runtime configurations, illustrating the trade-offs among chunk size, maximum delay, and computational complexity.

Future work will focus on further leveraging the linguistic knowledge in the pretrained decoder.

\newpage
\bibliographystyle{IEEEtran}
\bibliography{mybib}

\end{document}